\documentclass[twocolumn]{aastex62}

\usepackage{amsmath}
\graphicspath{{./}{figures/}}

\received{MONTH DD, YYYY}
\revised{MONTH DD, YYYY}
\accepted{MONTH DD, YYYY}
\submitjournal{ApJ}

\shorttitle{Seeing the unseen: a method to detect unresolved rings in protoplanetary disks}
\shortauthors{Scardoni et al.}

\begin{document}

\title{Seeing the unseen: a method to detect unresolved rings in protoplanetary disks}

\correspondingauthor{Chiara E. Scardoni}
\email{chiara.scardoni@unimi.it}

\author{Chiara E. Scardoni}
\affiliation{Institute of Astronomy, University of Cambridge, Madingley Road, Cambridge, CB3 0HA, UK}
\affiliation{Dipartimento di Fisica, Università degli Studi di Milano, Via Celoria 16, Milano, 20133, Italy}

\author{Richard A. Booth}
\affiliation{Astrophysics Group, Imperial College London, Blackett Laboratory, Prince Consort Road, London, SW7 2AZ, UK}
\affiliation{School of Physics and Astronomy, University of Leeds, Leeds, UK}

\author{Cathie J. Clarke}
\affiliation{Institute of Astronomy, University of Cambridge, Madingley Road, Cambridge, CB3 0HA, UK}

\author{Giovanni P. Rosotti}
\affiliation{Dipartimento di Fisica, Università degli Studi di Milano, Via Celoria 16, Milano, 20133, Italy}

\author{\'{A}lvaro Ribas}
\affiliation{Institute of Astronomy, University of Cambridge, Madingley Road, Cambridge, CB3 0HA, UK}



\begin{abstract}
{While high resolution ALMA observations reveal a wealth of substructure in protoplanetary discs, they remain incapable of resolving the types of small scale dust structures predicted, for example, by numerical simulations of the streaming instability.
In this Letter, we propose a method to find evidence for unresolved, optically thick dusty rings in protoplanetary disks. We demonstrate that, in presence of unresolved rings, the brightness of an inclined disc exhibits a distinctive emission peak at the minor axis. Furthermore, the azimuthal brightness depends on both the geometry of the rings and the dust optical properties; we can therefore use the azimuthal brightness variations to both detect unresolved rings and probe their properties. By analyzing the azimuthal brightness in the test-case of ring-like substructures formed by streaming instability, we show that the resulting peak is likely detectable by ALMA for typical disc parameters. Moreover, we present an analytic model that not only qualitatively but also quantitatively reproduces the peak found in the simulations, validating its applicability to infer the presence of unresolved rings in observations and characterize their optical properties and shape. This will contribute to the identification of disk regions where streaming instability (and thus planet formation) is occurring.}
\end{abstract}
f



\section{Introduction}
\label{sec:intro}
High resolution continuum observations with the Atacama Large Millimeter/sub-millimeter Array (ALMA) revealed a plethora of substructures in protoplanetary disks (\citealt{ALMA2015,Andrews2018}, see also the recent reviews by \citealt{Drazkowska+2023,Bae+2023}). As the data resolution increases, additional substructures may be uncovered \citep[see, e.g.,][]{Facchini+2020}, suggesting that, even at current high resolution, some substructures might not be resolved. Recent efforts focused on enhancing data analysis techniques; besides the standard deconvolution operated via the {\sc clean} algorithm \citep{Hogbom1974,Clark1980,Cornwell2008}, higher resolution techniques were developed, working both in the Fourier plane \citep[e.g.][]{Perkins+2015,Tazzari+2018,Jennings+2020} and in the image plane {\citep[e.g.][]{Akiyama+2017,Nakazato+2019,Zawadzki+2023}}. {Also some global disks observations suggest the presence of unresolved, optically thick substructures, e.g. the correlation between millimeter luminosity and the disk radius identified by \citet{Tripathi+2017} and \citet{Andrews+2018b}.} 
These works demonstrated that unresolved substructures may hide in existing data, and that not only the intrinsic data resolution, but also the subsequent data analysis, can help detailing further the disk dust distribution.

A very natural way to produce sub-au, optically thick substructures is dust filament formation via streaming instability (SI; \citealt{Youdin&Goodman2005,Johansen+2007,Johansen&Youdin2007,Youdin&Johansen2007}). SI is a powerful hydrodynamic instability driven by the dust-gas interaction; the difference in azimuthal velocity between the dust and the gas components gives rise to a drag force that slows down the dust grain motion, which thus loses some angular momentum and move to orbits closer to the central star (radial drift). The strongest radial drift is obtained for dust grains characterized by Stokes number $St\sim 1$ and results in the rapid loss of grains as they spiral toward the central star --  this process gives rise to an impediment to grain growth often referred to as the 'radial drift barrier' \citep{Adachi+1976,Weidenschilling1977,Takeuchi&Lin2002,Brauer+2008,Pinte&Laibe2014}. However, the dust backreaction can be strong enough to cause an increase in the gas azimuthal velocity (and thus a reduced radial drift) to the point of triggering SI. When triggered, SI promotes the formation of dusty dense filaments, tangential to the azimuthal direction, which can later collapse under the action of self-gravity and form planetesimals \citep{Johansen+2007,Youdin&Johansen2007,Carrera+2015,Yang+2017}. {SI is therefore at the core of our planet formation theories, but the SI-induced substructures are too small ($\lambda\ll H$) to be directly resolved with our current observations. Indirect techniques to study this mechanism can thus greatly aid in characterizing this effect and its impact in the planet formation process.}

In this context, we aim at exploring whether unresolved substructures can leave a detectable imprint in the observations. We specifically focus on dense, optically thick rings over an optically thin background, and analyze the azimuthal brightness variation in this configuration. Due to the combined effect of the rings' geometry and optical depth, we expect maximum (minimum) emission at the minor (major) axis, when the disk is observed at an inclination; we develop a qualitative understanding and simplified theoretical model of such emission in \sectionautorefname~\ref{sec:AnalyticalModel}. In \sectionautorefname~\ref{sec:SIsims} we present simulations of SI, and apply our model to the resulting tangential substructures, confirming that the model successfully reproduces the emission obtained from the simulations. We further show that, under realistic physical conditions, the difference between the major and minor axis is higher than the $10\%$; this implies that unresolved tangential substructures are potentially detectable in the disk, even if their size is below ALMA resolution limits. Our findings are then summarized in \sectionautorefname~\ref{sec:Conclusions}.

\subsection{Azimuthal brightness asymmetries}
\section{{Toy model}}
\label{sec:AnalyticalModel}
\begin{figure*} 
\centering
\includegraphics[width=0.9\textwidth]{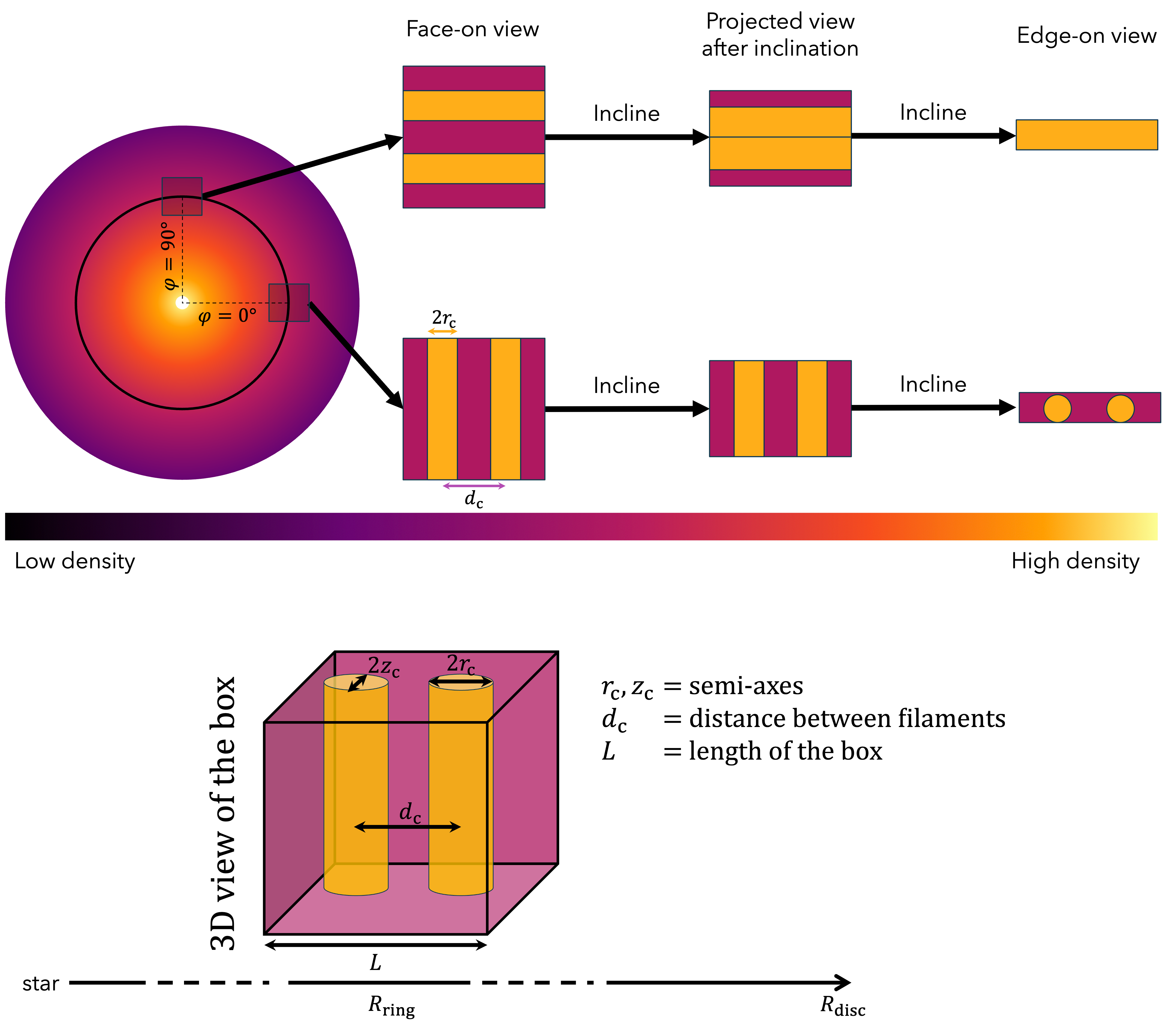}
\caption{{Upper panel:} sketch of a protoplanetary disk containing two unresolved dense rings. The drawings on the right illustrate the local view of the unresolved rings at $\varphi=0^{\circ}$ and $\varphi=90^{\circ}$ for a face-on (boxes on the left), inclined (center), edge-on (right) disk. {Lower panel: 3D view of the box showing the model key parameters (as an example, we consider the cylinder model).}}
\label{fig:EmissionScheme}
\end{figure*}
As an illustration we consider an apparently smooth disk and assume that it contains two unresolved, optically thick cylindrical rings with a finite vertical extent at a given radial location $R$ (see the sketch in \figureautorefname~\ref{fig:EmissionScheme}); at any azimuthal location we therefore expect to measure the emission from both the optically thick rings and the optically thin background, averaged over the observable area. To introduce a qualitative understanding of the azimuthal profile of the disk emission, we focus on two disk patches located at $\varphi=0^{\circ}$ and  $\varphi=90^{\circ}$ {(corresponding to the major and minor axis, respectively)} and analyze their emission in both the face-on and inclined case, as illustrated in \figureautorefname~\ref{fig:EmissionScheme} (in the sketch the rings are locally represented by the yellow bands).

For a face-on disk, at different azimuthal angles the local orientation of the rings varies within the observable area; the relative area of the disk patch occupied by the optically thick substructures, however, remains the same at all azimuthal angles. This implies that the averaged emission is independent of the azimuthal angle.

If we instead observe the disk at an inclination, the projected lengths along the vertical axis (i.e. the minor axis) are shortened with respect to the face-on case. At $\varphi=0^{\circ}$ this does not impact the relative area of the disk patch occupied by the rings, as both the rings and the background projection are subject to the same shortening effect. Here, the inclination only has the effect of increasing the column density, leading to an increase of the overall optical depth.
The increase in the column density is also present at $\varphi=90^{\circ}$. {On top of that the shortening of the projected lengths along the vertical axis determines the reduction of the optically thin area between the two rings; this happens because the optically thick rings have a finite thickness which causes the optically thick areas to infill the optically thin areas when the inclination is sufficiently large.} 
This effectively increases the relative area occupied by the optically thick rings. {In the limiting case of a edge-on disk, we thus expect the emission at  $\varphi=90^{\circ}$ to be completely optically thick (unless the rings are infinitesimally geometrically thin).} As a consequence the local average optical depth is higher than that at $\varphi=0^{\circ}$. 
The overall effect is that the disk observed at an inclination is characterized by higher (lower) emission at the minor (major) axis.

\subsection{Clump covering factor}
We model the effect illustrated above by defining a local box associated with each azimuthal location and computing the averaged emission within the box. Each box contains an optically thin background and a number of optically thick elongated substructures ('walls'), resembling  resembling sections of unresolved rings. We consider two models for the geometry of the substructures: cylinders with elliptic cross-section (semi-axes $r_{\rm c}$ and $z_{\rm c}$) and cuboids with rectangular cross-section (semi-sides $r_{\rm r}$ and $z_{\rm r}$).\footnote{In the rest of the Letter, we use the subscript `c'/`r' to refer to the cylinders/rectangular cuboids model. Note that in both cases $r$ represents half of the face-on width of the substructures.} If $r_{\rm c}=r_{\rm r}$ these models are equivalent when the disk is face-on, but their 3D structure determines distinct effects when observed at an inclination. {We define as $d_{\rm c,r}$ the distance between the projected center of two consecutive substructures and $L$ as the face-on width of the box, thus the number of rings contained in each box is given by $N_{\rm c,r}=L/d_{\rm c,r}$.}

For a given inclination angle $i$, the projection on the sky of the box area is $A(i) = A_{0} \cos(i)$, where $A_{0}$ is the face-on box area. 
When the substructures are modeled as cylinders, their area varies with $\varphi$ and $i$ as follows
\begin{equation}
    A_{\rm c}=2(N_{\rm c}-1)h_{\rm c,0}\sqrt{r_{\rm c}^2\cos^2 i + z_{\rm c}^2\sin^2 i \sin^2\varphi},
    \label{eq:cylArea}
\end{equation}
where $h_{\rm c,0}$ is the face-on cylinder height. In the rectangular cuboids model, the area occupied by substructures is

\begin{equation}
\begin{aligned}
    A_{\rm r}=&2(N_{\rm r}-1)h_{\rm r,0}\sqrt{\cos^2 \varphi \cos^2 i +\sin^2 \varphi}\cdot \\& \cdot \left(r_{\rm r}\sqrt{\cos^2 \varphi +\cos^2 i \sin^2\varphi}+z_{\rm r}\sin i |\sin\varphi|\right).
    \label{eq:cubArea}
\end{aligned}
\end{equation}
{The area of the background is then determined as $A_{\rm bg}=A_{\rm box}-A_{\rm c,r}$.}
The rings' covering factor is given by the substructure-to-box area ratio $A_{\rm c,r}/A_{\rm box}$; this parameter determines the portion of the box occupied by the rings and, therefore, it regulates the average optical depth at any $\varphi$.

\subsection{Optical depth and averaged emission}
To find the average emission at each $\varphi$, we start by associating a face-on value for the optical depth to both the substructures ($\tau_{\nu;\rm c,r}^{0}$) and the background ($\tau_{\nu,\rm bg}^{0}$). The optical depth of the background for the inclined system is 
\begin{equation}
    \tau_{\nu,\rm bg} = \frac{\tau_{\nu,\rm bg}^{0}}{\cos i},
\end{equation}
The optical depth of the inclined substructures is given by
\begin{equation}
    \tau_{\nu;\rm c,r} = \tau_{\nu;\rm c,r}^{0}\frac{r_{\rm c,r}}{ \sqrt{ r_{\rm c,r}^2\cos^2 i + z_{\rm c,r}^2 \sin^2 i \sin^2\varphi}},
\end{equation}
where we account for the variations in column density at different azimuthal locations.\footnote{We assume that the substructures are sufficiently optically thick that we do not need to consider variations of optical depth within them.} However, as we will show later, as long as the substructures are optically thick ($\tau_{\nu;\rm c,r}^0>5$), the exact value of $\tau_{\nu;\rm c,r}$ is unimportant for the purposes of this Letter.

By defining a temperature profile {as in \citet{Tazzari+2020a,Tazzari+2020b}}
\begin{equation}
    T(R) = 120 {\rm K} \left(\frac{R}{\rm au}\right)^{-3/7},
    \label{eq:temperatureprofile}
\end{equation}
we can find the emission of both the cylinders and the background as
\begin{equation}
    I_{\nu} = B_{\nu}(T)(1-{\rm e}^{-\tau_{\nu}}),
\end{equation}
where $B_{\nu}(T)$ is the Plank function. We finally obtain the observable emission as the emission averaged over the box (i.e. the observable area)
\begin{equation}
    \langle I_{\nu}\rangle_{\rm box} = \frac{I_{\nu;\rm c,r} A_{\rm c,r} + I_{\nu,\rm bg} A_{\rm bg}}{ A_{\rm box}}.
    \label{eq:Inu_average}
\end{equation}

\subsection{Impact of the model parameters on the azimuthal brightness}
\label{sec:ModelParameters}
\begin{figure*}
\centering
\includegraphics[width=1\linewidth]{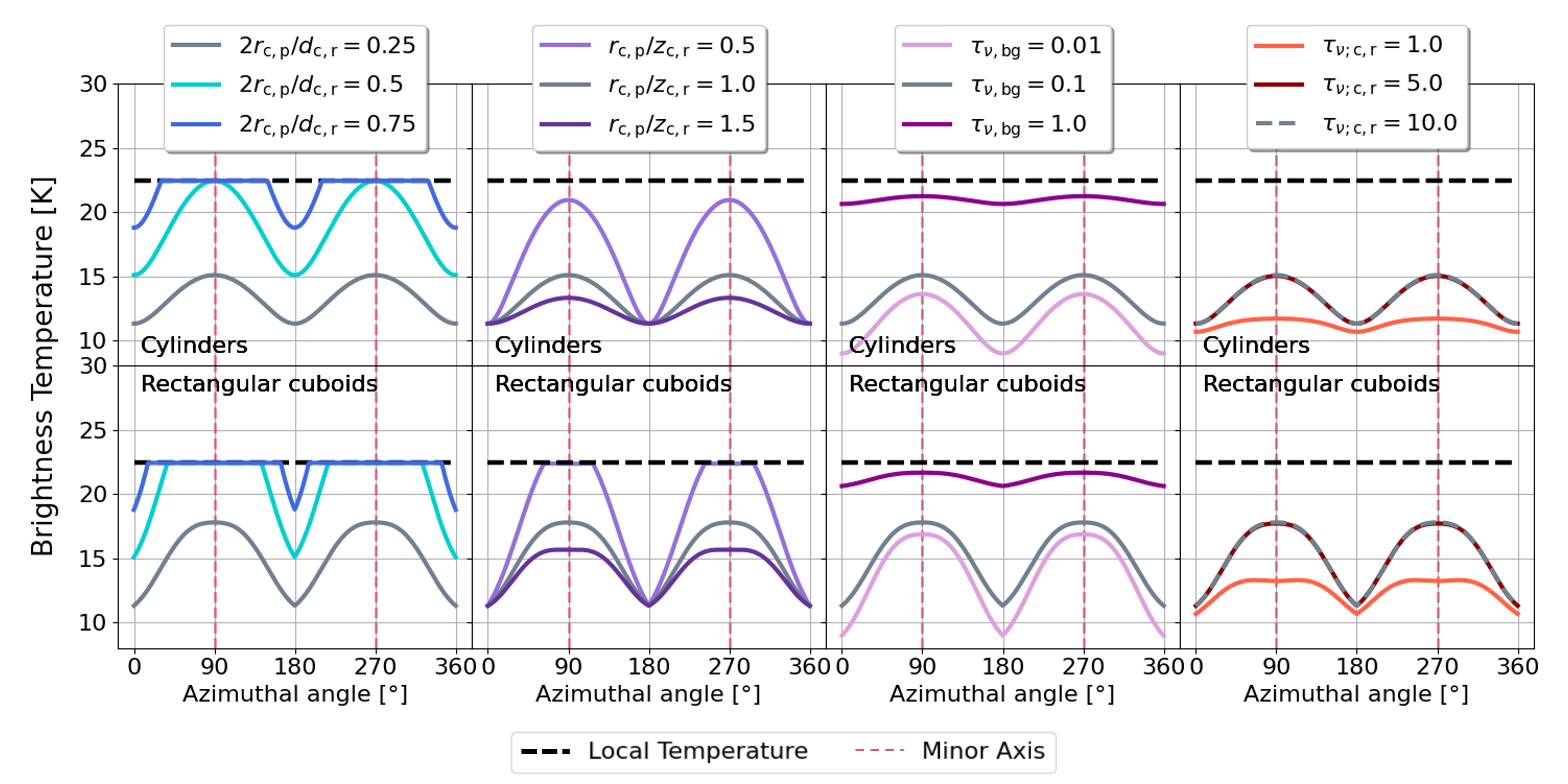}
\caption{Brightness temperature at $50\ \rm au$ as a function of the azimuthal angle for disks with $60^{\circ}$ inclination containing unresolved, optically thick rings. The plots in the upper (lower) row refer to the case where rings are locally modeled as cylinders (rectangular cuboids). In each column the effect of varying a given system parameter (indicated in the legend) is analyzed. The black dashed line corresponds to the local temperature, {while the vertical red dashed lines indicate the location of the minor axis.}}
\label{fig:BrightnessTparameters}
\end{figure*}
Our model comprises 4 parameters, 2 geometric parameters ($2r_{\rm c}/d_{\rm c}$, $r_{\rm c}/z_{\rm c}$) and 2 optical parameters ($\tau_{\nu,\rm c}$, $\tau_{\nu,\rm bg}$). {{Note that $2r_{\rm c}/d_{\rm c}$ is limited by $0\leq 2r_{\rm c}/d_{\rm c}\leq 1$, with $0$ ($1$) corresponding to the case where the rings occupy no (all) portion of the emitting area.}} Additionally, we must define a temperature profile (\equationautorefname~\ref{eq:temperatureprofile}) and the disk inclination.

We analyze the impact of the model parameters on the brightness temperature azimuthal profile (which probes the azimuthal emission of the disk)
\begin{equation}
    T_{\rm B} = \frac{h\nu}{k_{\rm B}}\left[\log\left(1+\frac{2h\nu^3}{c^2\langle I_{\nu}\rangle_{\rm box}}\right)\right],
\end{equation}
where $h$ and $k_{\rm B}$ are the Plank and Boltzmann constants, respectively; $\nu$ is the frequency of the observation (we use the frequency corresponding to ALMA band 6); $c$ is the light speed. The parameter analysis is shown in \figureautorefname~\ref{fig:BrightnessTparameters}, when the dense rings are locally modeled as cylinders (upper row) and rectangular cuboids (lower row). We take $2r_{\rm c}/d_{\rm c}=0.25$, $r_{\rm c}/z_{\rm c}=1$, $\tau_{\nu,\rm bg}=0.1$, $\tau_{\nu,\rm c}=10$ as reference parameters (gray curve in all the plots), and compute the emission at the reference radius $50\ \rm au$. The black dashed line shows the local temperature of the disk. In each panel we show the effect of varying one of the parameters, as indicated in the legend above each panel column. In the following we will refer to the brightness temperature at the minor (major) axis as $T_{\rm min}$ ($T_{\rm maj}$).

The parameter $2r_{\rm c}/d_{\rm c}$ (first column) influences the covering factor of the rings, and therefore the extent of the optically thick portion of the emitting area. The higher is $2r_{\rm c}/d_{\rm c}$, the bigger is the portion occupied by the optically thick substructures and, thus, $T_{\rm maj}$. In some cases the brightness temperature saturates at the local temperature; this happens when $2r_{\rm c}/d_{\rm c}$ is high enough to have the substructures covering the entire emitting area for some azimuthal locations and it is favored by high $2r_{\rm c}/d_{\rm c}$.

The parameter $r_{\rm c}/z_{\rm c}$ (second column) determines the geometry of the rings' section; high (low) values of the parameter correspond to substructures wider (taller) than tall (wide). {As noted previously lower $r_{\rm c}/z_{\rm c}$ (narrow tall rings) lead to a more pronounced increase in optical depth along the minor axis.}  When the disc is inclined, indeed, the substructure thickness contributes to the projected area, with the biggest contribution given at $\varphi=90^{\circ}$; this implies that for lower values of $r_{\rm c}/z_{\rm c}$ the increase of optical depth at the minor axis (and thus $T_{\rm min}$) is more pronounced. On the other hand, $T_{\rm maj}$ is independent of $r_{\rm c}/z_{\rm c}$ because at that location we can only see the width of the substructure, regardless of the disc inclination.

We now consider the optical parameters (third and fourth columns). As evident from the plots in the third column, the value of $\tau_{\nu, \rm bg}$ determines $T_{\rm maj}$, with higher $\tau_{\nu, \rm bg}$ corresponding to higher values of $T_{\rm maj}$ (this is because the emission is overall optically thicker in this case). {Provided that the rings are optically thick ($\tau_{\nu, \rm c}>1$) and the background is optically thin ($\tau_{\nu, \rm bg}<1$), the contrast between the brightness temperature at the minor and major axis is roughly independent of {$\tau_{\nu, \rm bg}$}. Since we are considering optically thick rings ($\tau_{\nu, \rm c}>1$) {$\tau_{\nu, \rm c}$} can be discarded, leaving us with a 3 parameter model.}

Finally, by comparing the upper and lower row we can notice that modeling the substructures as cylinders produces a $T_{\rm B}$ oscillation smaller than the rectangular cuboids' case; furthermore, the curves associated with the rectangular cuboids tend to show a flattening in the area around the minor axis. Both these differences are due to the different contribution given by the component of the substructures that become visible when the system is inclined (see \equationautorefname~\ref{eq:cylArea} and \equationautorefname~\ref{eq:cubArea}), which determines a sharper transition in the case of the rectangle (since it is characterized by $90^{\circ}$ angles). In real cases, we can expect an intermediate behavior between the two illustrated cases; with the $T_{\rm B}(\varphi)$ curve determined by the specific 3D shape of substructures.

\section{Streaming instability}
Since SI is a natural way to obtain unresolved, optically thick substructures, we use SI-induced substructures as a case-study for our model.

\subsection{Local simulations of streaming instability}
\label{sec:SIsims}
\begin{figure} 
\centering
\includegraphics[width=1\linewidth]{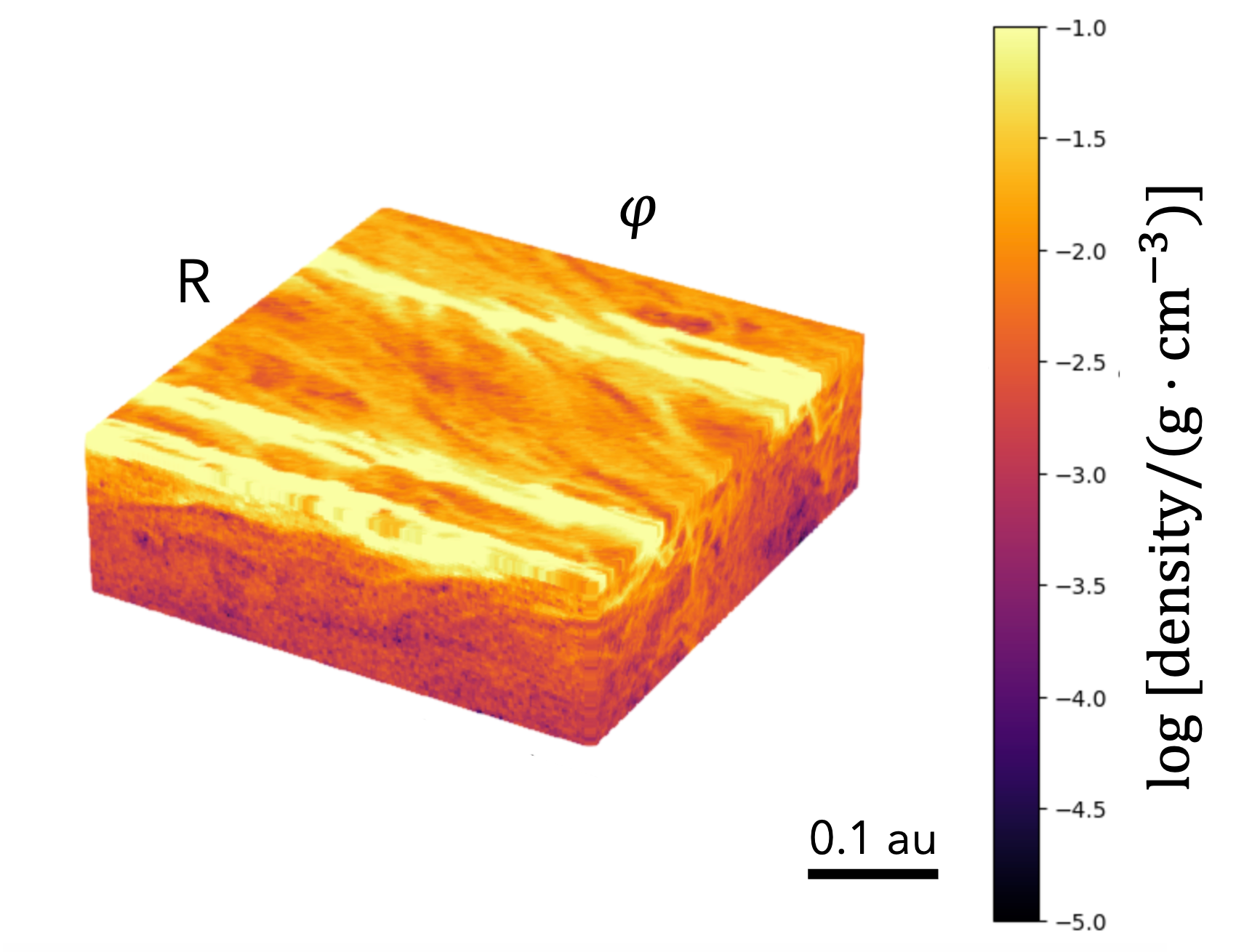}
\caption{Dust density obtained from a 3D local simulation of a system undergoing SI. The yellow overdense filaments are the SI induced dust clumps.}
\label{fig:SteamingInstability3D}
\end{figure}

{To analyze systems undergoing SI we employ the 2D (radial and vertical directions) local simulations\footnote{{Global SI simulations are computationally prohibitive; however, previous studies \citep[e.g.][]{Yang&Johansen2014,Li+2018} demonstrated that dust concentration in filaments in the non-linear phase of SI remains robust for bigger boxes.}}} {presented in \citet{Scardoni+2021}, performed with the hydro-code {\sc Athena} \citep{Bai&Stone2010,Bai&Stone2010-2,Bai&Stone2010_athenaparticles}. The simulated boxes' size is $L_r\times L_z=0.05\times 0.15$, with resolution $N_r\times N_z=256\times 768$. We consider $N_{\rm tot}=7\cdot 10^5$ particles (equally distributed among 28 Stokes numbers) for which we adopt (shear) periodic boundary conditions in the radial direction, and reflecting boundary conditions in the vertical direction. We refer to \citet{Scardoni+2021} for details about the simulation set; in the following we show results for the simulation characterized by: local dust-to-gas mass ratio $Z=0.03$; particle number distribution $n(a)=n_0a^{-3.5}$; pressure support parameter $\Pi=0.025$; Stokes numbers between $\tau_{\rm s}=10^{-4}-10^{-1}$. The parameters choice relies on previous studies where the combinations of parameters triggering SI were identified \citep[e.g.][]{Bai&Stone2010-2,Carrera+2015,Johansen+2006,Johansen+2007}. }

{To analyze the azimuthal emission properties, we need the 3D dust distribution. In principle, the azimuthal direction is unimportant as all the parameters regulating SI only depend on the radius; therefore, we built 3D boxes from the 2D simulations by assuming that the SI-induced substructures are azimuthally symmetric. Practically, the 2D boxes are converted into 3D boxes by replicating the 2D surface density in $R-z$ for $N_{\varphi}=256$ times along the $\varphi$ direction; this is equivalent to actual 3D boxes with perfectly symmetric azimuthal evolution.}

{To ensure that the 3D-equivalent simulations are appropriate for our purposes, we perform two 3D test-simulations, characterized by box of size $L_r\times L_{\varphi}\times L_z=0.05\times 0.05\times 0.15$ and resolution $N_r\times N_{\varphi}\times N_z=256\times 256\times 768$. We consider the following parameters: $Z=0.03$, $n(a)=n_0a^{-3.5}$, $\Pi=0.025$, $\tau_{\rm s}=10^{-4}-10^{-1}$ (first simulation) and $\tau_{\rm s}=10^{-3}-1$ (second simulation). \figureautorefname~\ref{fig:SteamingInstability3D} illustrates the typical substructures obtained in the 3D simulations (converted in physical units, see \sectionautorefname~\ref{sec:physandradtransf}): the dust filaments are dense, sub-au, tangential substructures, thus they have the same properties as the system modeled in \sectionautorefname~\ref{sec:AnalyticalModel}; we also notice that the SI-induced substructures are approximately azimuthally symmetric.
Due to the high computational cost, however, these simulations only run for $300\ \Omega^{-1}$; by that timestep dust filaments have been formed, but their properties have not fully reached steady state, which is expected to require $\sim1500-2000\ \Omega^{-1}$ for this choice of parameters \citep{Scardoni+2021}. For this reason in the Letter we use the 3D-equivalent simulations and demonstrate the appropriateness of this choice in the right panel of \figureautorefname~\ref{fig:testModel}.}

\subsection{Physical and radiative transfer model}
\label{sec:physandradtransf}
To analyze the emission from the simulated systems, we need to convert the dimensionless simulations to physical units; for this purpose we take as reference model the MMSN in \citet{Chiang&Youdin2010}. The length unit is defined by the local gas height {$ H_{\rm g}(R)=h_0 R\left({R}/{\rm au}\right)^{2/7}$, where $h_0=0.04$ is the disk aspect ratio at $1\ \rm au$.}
The mass unit is defined by the local gas density $\Sigma_{\rm g}(R)=2200\ \rm g/cm^2\ \left({R}/{\rm au}\right)^{-1.5}$.

We define an integrated disk model extending from $1-100\ \rm au$, and split it into $N_R\times N_{\varphi}=250\times 1000$ cells. We apply the local physical units to map the simulated box to each cell, and `build' the disk, then we incline it by a given angle (we take $i=60^{\circ}$) and compute its emission at all $(R,\varphi)$ coordinates.

To find the locally averaged emission from the $(R,\varphi)$ locations, we adopt a simple analytic radiative transfer model. Within the simulated box, we compute the optical depth as
\begin{equation}
	\tau_{\nu}=\frac{k_{\nu}^{\rm avg}\Sigma_{\rm d}}{\cos{i}},
	\label{eq:optDepth}
\end{equation}
where $k_{\nu}^{\rm avg}$ is the averaged opacity (which includes information from all the particles included in the simulation),\footnote{We use \citealt{Birnstiel2018} code, assuming spherical compact grains with the following composition (and volume fractions): water (0.6), silicates (0.1) \citep{Warren&Brandt2008,Draine2003} and organics (0.3) \citep{Zubko+1996}.} $\Sigma_{\rm d}$ is the dust density, $i$ is the disk inclination angle. We then obtain the specific intensity within the box as
\begin{equation}
    I_{\nu}=B_{\nu}(T)(1-{\rm e}^{-\tau_{\nu}}),
    \label{eq:SpecInt}
\end{equation}
and we average this throughout the simulated box as in \equationautorefname~\ref{eq:Inu_average}, obtaining the observable emission.

\subsection{Analysis of the azimuthal brightness}
\begin{figure*} 
\centering
\includegraphics[width=1\linewidth]{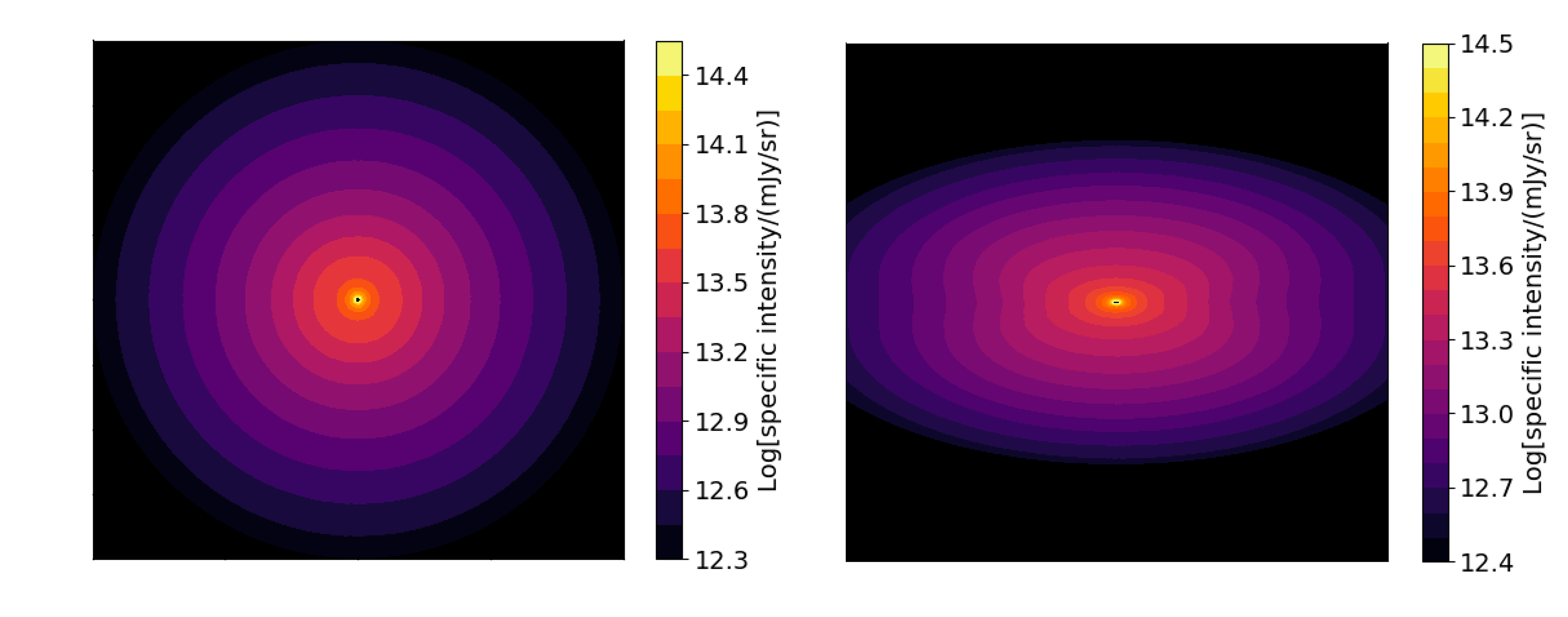}
\caption{2D maps of a protoplanetary disk undergoing SI obtained by mapping the local simulation to different locations in the disk. On the left we show the face-on view, on the right the same disk is seen at $60^{\circ}$ inclination.}
\label{fig:2Dmaps}
\end{figure*}
In \figureautorefname~\ref{fig:2Dmaps} we show the 2D emission maps at ALMA band 6 of the simulation described in \sectionautorefname~\ref{sec:SIsims}. As expected, the face-on system (plot on the left) is azimuthally symmetric; {looking at the isophotes of the inclined system (on the right), we can notice that they are not elliptical (as expected for a smooth disc), but curve towards smaller radii along
the major axis, indicating the low brightness temperature on that axis.}

As a preliminary test of detectability, we analyze the $T_{\rm B}$ profiles overall the disk, and select the radii where $\Delta T > 10\% \langle T_{\rm B}\rangle$ (where $\Delta T = |T_{\rm min}-T_{\rm maj}|$) -- {we chose this threshold because the amplitude of azimuthal variations measured by \citet{Doi&Kataoka2021} is $10\%$, showing that a deviation of this kind would be detectable}. In the considered simulation, this condition is satisfied for radii from $\sim 25-65\ \rm au$. {For smaller radii, the high local density makes the background optically thick, reducing the contrast between $\tau_{\nu,\rm bg}$ and $\tau_{\nu,\rm c}$, and thus reducing the $T_{\rm B}$ oscillation. Conversely, at larger radii the local density is lower and the filaments become optically thinner, thus the contrast between $\tau_{\nu,\rm bg}$ and $\tau_{\nu,\rm c}$ is reduced and the $T_{\rm B}$ oscillation is milder (see \appendixautorefname~\ref{a:radvariation}).} This analysis confirms that, for typical disk parameters, unresolved optically thick rings would be detectable at a wide range of radii.

\begin{figure*}
\centering
\includegraphics[width=0.52\linewidth]{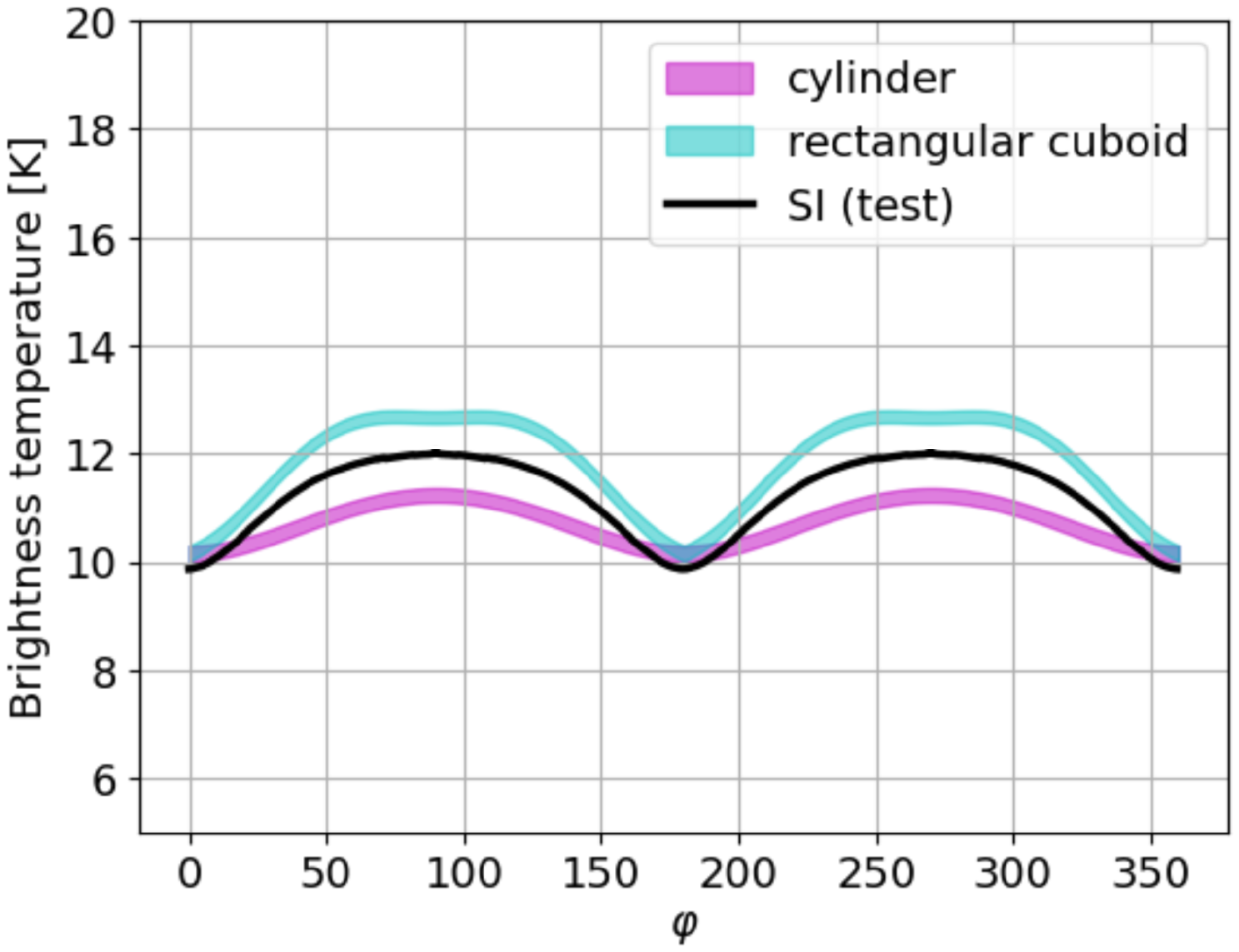}%
\includegraphics[width=0.48\linewidth]{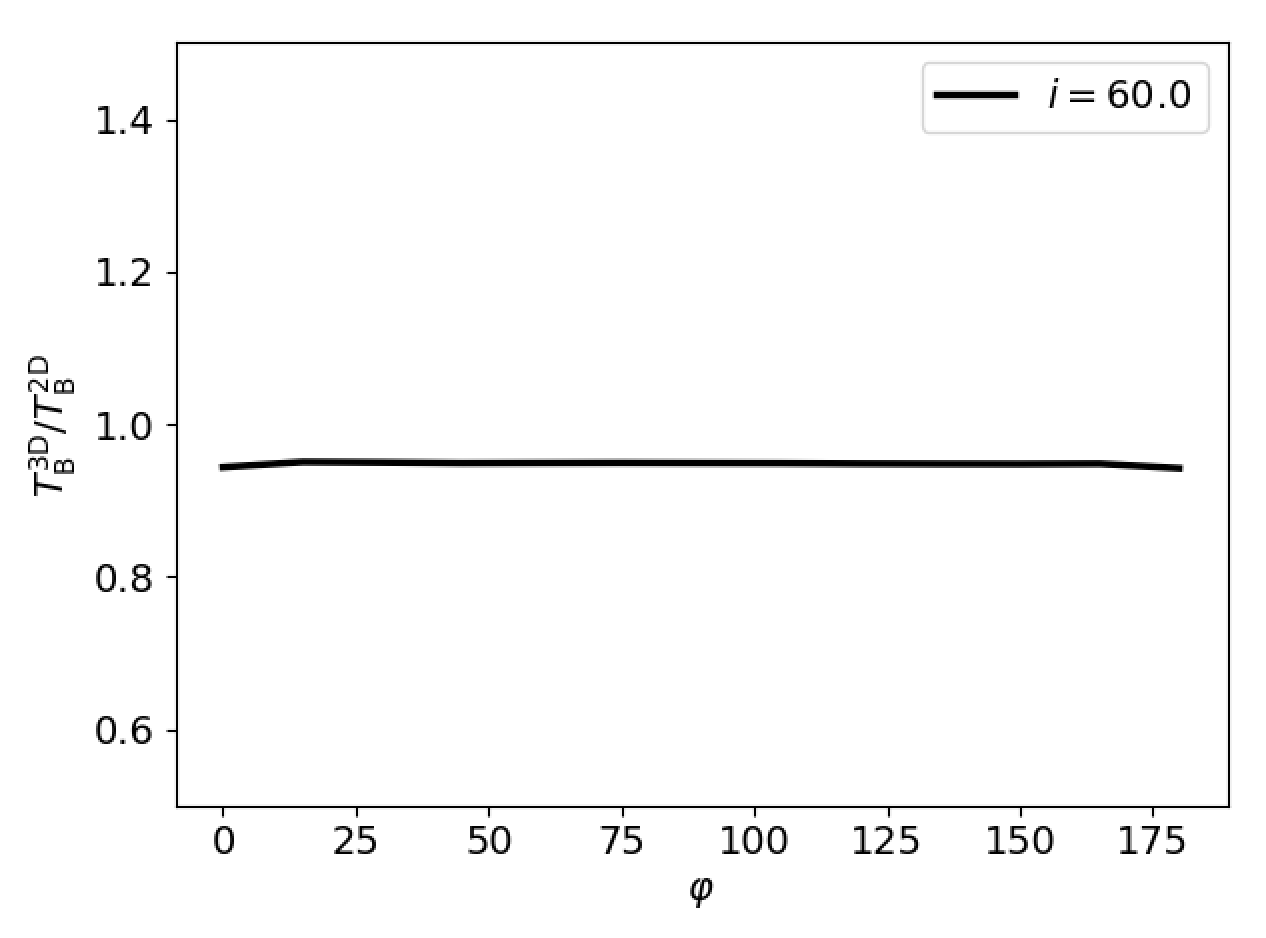}
\caption{{Left panel: brightness temperature azimuthal profile at 50 au for the SI simulation (black line); the magenta and cyan lines are obtained from our model in the cylinder and rectangular cuboid configurations, respectively. Right panel: ratio between the brightness temperature between the 3D simulations and the 3D-equivalent version of the 2D simulations, considering a disc inclination of $60^{\circ}$.}}
\label{fig:testModel}
\end{figure*}
We select an arbitrary reference radius within the range where the $\Delta T > 10\% \langle T_{\rm B}\rangle$, namely $R=50\ \rm au$, to test whether our analytic model can reproduce the simulated emission. For this purpose, we estimate from the simulation the values for the 3 model parameters: the geometric parameters {$2r_{\rm c}/d_{\rm c}\sim 0.2$ and $r_{\rm c}/z_{\rm c}\sim 0.33$ are found by analyzing the dust density distribution in the simulation; the background optical depth $\tau_{\nu, \rm bg}\sim 0.1$} is determined by taking the opacity model and finding the value corresponding to the size of the group of small grains contained in the simulation (we consider a $10\%$ error on the estimated $\tau_{\nu, \rm bg}$ value). Since we demonstrated that the filaments' optical depth is unimportant as long as they are optically thick, we arbitrarily take $\tau_{\nu, \rm c}=10$. The azimuthal brightness temperature profiles are shown in \figureautorefname~\ref{fig:testModel}, where the magenta/cyan lines are the results for the cylinder/rectangular cuboids model, while the black line shows the brightness temperature resulting from the simulations at the reference radius. The simulation results are located between the two models, potentially implying that the specific shape of the filaments is more complex than that modeled; indeed, from \figureautorefname~\ref{fig:SteamingInstability3D}, we see that the filaments are not perfectly represented by either considered model. 
{This suggests that the $T_{\rm B}-\varphi$ curve shape can be used to constrain the shape of the unresolved rings.}

{We caution that in the model we do not account for the observation resolution. To test the resolution effect, we convolved the image with a Gaussian beam with ${\rm FWHM}= 15\ \rm au$ (resolution of $0.1$ arcsec at $\sim 150\ \rm pc$), finding that the $T_{\rm B}$ oscillation is reduced of $\sim 0.2$ K. This suggests that the signature is still observable for favorable conditions of the system.}

{Finally, we test that using the 3D-equivalent simulations is appropriate for our purposes by computing the ratio between the brightness temperature of the 3D simulations and the 3D-equivalent simulations (right panel of \figureautorefname~\ref{fig:testModel}), computed at $300\ \Omega^{-1}$. Since the difference between the two simulations is $\sim 5\%$, we confirm that the 3D-equivalent simulations are a good approximation of the 3D evolution of the system.}

\section{Discussion and conclusion}
\label{sec:Conclusions}
In this Letter we proposed a new method to detect unresolved, optically thick rings. As illustrated in \sectionautorefname~\ref{sec:AnalyticalModel}, when the disk is inclined, the specific combination of geometry and optical properties of the rings causes a characteristic emission peak at the minor axis, and we argue that this effect is uniquely related to unresolved optically thick rings. Indeed, when all the system is optically thick, we expect $T_{\rm B}$ to be constant and equal to the local temperature; conversely, \citet{Doi&Kataoka2021} demonstrated the opposite effect in the case of optically thin (but geometrically thick) rings, where the emission is stronger at the major axis. No other azimuthal emission asymmetries analyzed so far found the same effect as that presented in this Letter; for example, \citet{D'Alessio+2004} suggested that a wall (i.e. a location where the emission transits from optically thin to optically thick) in the emission might determine a maximum in the emission at $\varphi=90^{\circ}$ and a minimum at $\varphi=270^{\circ}$. {The shadowing from a misaligned inner disk might produce a similar signature for very specific system's parameters \citep{Min+2017}, this would however require the shadow to be aligned with the major axis (which depends on the orientation of the disc with respect to the line of sight). This event is a random chance and would therefore require a fine tuning of the parameters, we thus expect this situation to be unlikely.
If the predicted signature is found in the observations, we can check whether the effect is caused by a misaligned inner disk by using information from GRAVITY observations \citep[see, e.g.,][]{Bohn+2022}.}
The peak (minimum) at the minor (major) axis is therefore peculiar to optically thin systems containing annular optically thick substructures {(except for specific misaligned inner disks)} and, therefore, can be used as a detection method for such structures.

We demonstrated that the parameters regulating the brightness temperature profiles are the ring size-to-distance ratio $2r/d$, the ring width-to-height ratio $r/z$ and the background optical depth $\tau_{\nu,bg}$ (see \sectionautorefname~\ref{sec:ModelParameters}). We linked these parameters to the 3 main observables
\begin{enumerate}
    \item The brightness temperature oscillation between the minor and major axis $\Delta T$ is primarily dependent of $2r/d$ and $r/z$, with $\Delta T$ increasing for higher $2r/d$ and lower $r/z$.
    \item The brightness temperature at the major axis $T_{\rm maj}$ is mainly influenced by $\tau_{\rm bg}$ and $2r/d$, with higher $T_{\rm maj}$ for higher values of these parameters.
    \item The detailed shape of the azimuthal profile is dependent on the 3D geometry of the ring (i.e. whether the ring cross-section is more elliptical or rectangular).
\end{enumerate}
Therefore, if we have an independent measure of the disk inclination, we can gain information on the dust optical depth and the 3D structure of the unresolved rings by fitting the brightness temperature azimuthal profile.

This method is thus a valuable tool to search for observational signatures of sub-au dusty ring-like substructure, which would otherwise remain undetected. We demonstrated the applicability of our model to systems undergoing SI (expected to form dense, elongated, sub-au substructures), showing that: {(i) for typical disc parameters, the brightness temperature oscillation is $>10\%$, thus likely detectable by ALMA (\citealt{Doi&Kataoka2021} measured $T_{\rm B}$ variations of $\sim 10\%$)}; (ii) the brightness temperature profile from simulations is successfully reproduced by the model. This confirms the model capability to detect and characterize substructures formed in the intermediate stages of planet formation; more generally, we expect our model to be useful to test all the processes that predict the formation of elongated dust substructures.

{We highlight that we neglected the scattering, whose effect would make the optically thick substructures to appear fainter, determining a lower $\Delta T$. However, the effect is likely small: for optically thick emission ($\tau_{\nu,\rm c} \geq 10$), we would need albedo $> 0.8$ to have $I_{\nu}/B_{\nu} <0.5$ \citep{Zhu+2019}. {Thus, although a factor $\sim 2$ reduction in $\Delta T$ might happen, it is unlikely to have a complete removal of the signature due to scattering.}}

{We finally caution that in this Letter we considered symmetric optically thick rings in otherwise smooth disks. Even in this simplified case, it is relatively hard to see the signature `by eye' from the 2D maps (see \figureautorefname~\ref{fig:2Dmaps}). We thus anticipate that, when considering real observations, only a careful analysis of $T_{\rm B}-\varphi$ profiles may reveal the peak at the minor axis. If the signature is found in observations, the model can probe the unresolved substructures; otherwise, it can still provide upper limits on the properties of any unresolved substructure that might be present.}


\begin{acknowledgments}
We thank the anonymous referee for the helpful comments and suggestions that improved the quality of this letter. 
CES acknoledges support from Peterhouse Cambridge and the Institute of Astronomy (University of Cambridge).
CES and GPR acknowledge support from the European Union (ERC Starting Grant DiscEvol, project number 101039651) and from Fondazione Cariplo, grant No. 2022-1217. Views and opinions expressed are, however, those of the author(s) only and do not necessarily reflect those of the European Union or the European Research Council. Neither the European Union nor the granting authority can be held responsible for them. 
RAB acknowledges support from the Royal Society in the form of a University Research Fellowship.
AR has been supported by the UK Science and Technology research Council (STFC) via the consolidated grant ST/W000997/1.
This work has also been supported by the European Union’s Horizon 2020 research and innovation programme under the Marie Sklodowska-Curie grant agreement number 823823 (DUSTBUSTERS).
This work was in part performed using the Cambridge Service for Data Driven Discovery (CSD3), part of which is operated by the University of Cambridge Research Computing on behalf of the STFC DiRAC HPC Facility (www.dirac.ac.uk). The DiRAC component of CSD3 was funded by BEIS capital funding via STFC capital grants ST/P002307/1 and ST/R002452/1, and STFC operations grant ST/R00689X/1.
\end{acknowledgments}

%

\vspace{5mm}





\appendix
\section{Results at different radii}
\label{a:radvariation}
\begin{figure*} 
\centering
\includegraphics[width=0.5\linewidth]{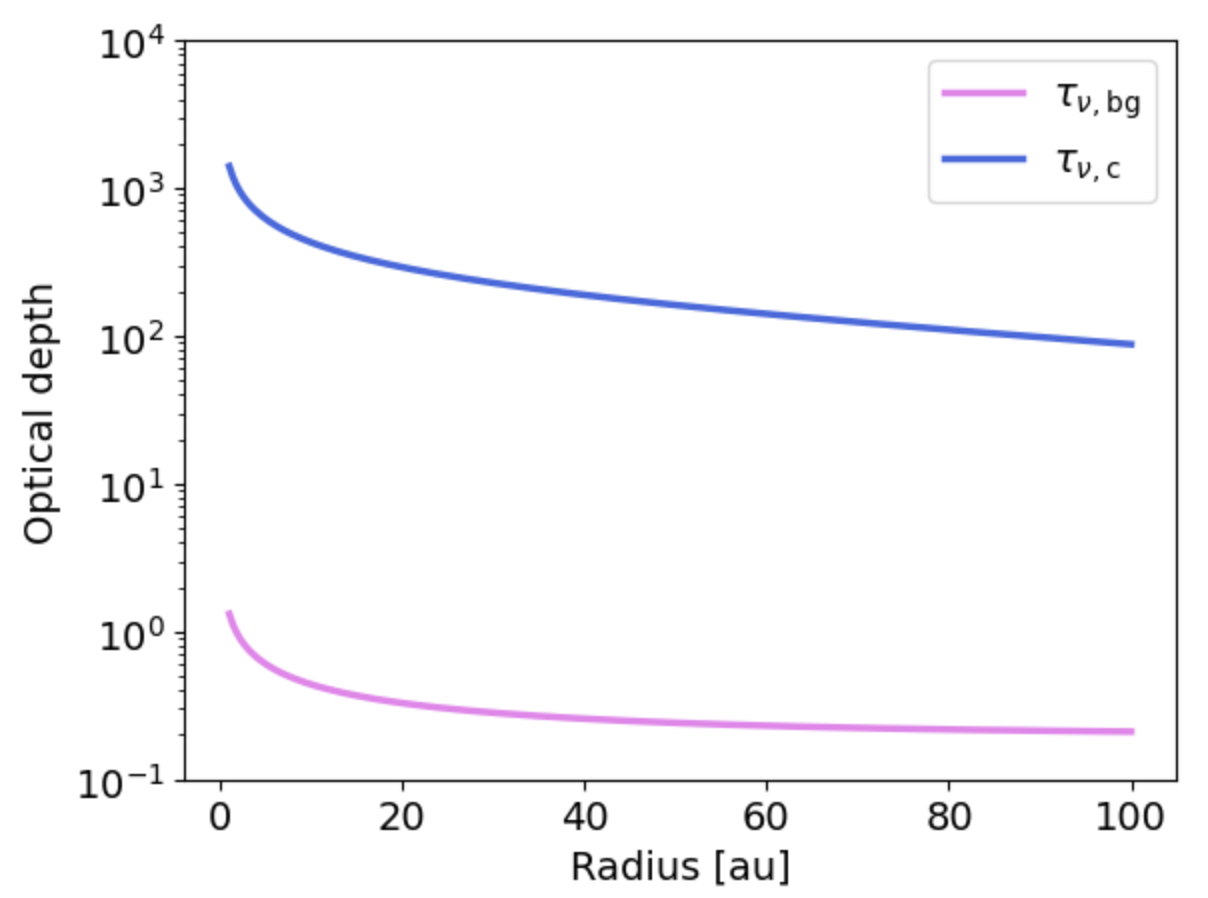}%
\includegraphics[width=0.5\linewidth]{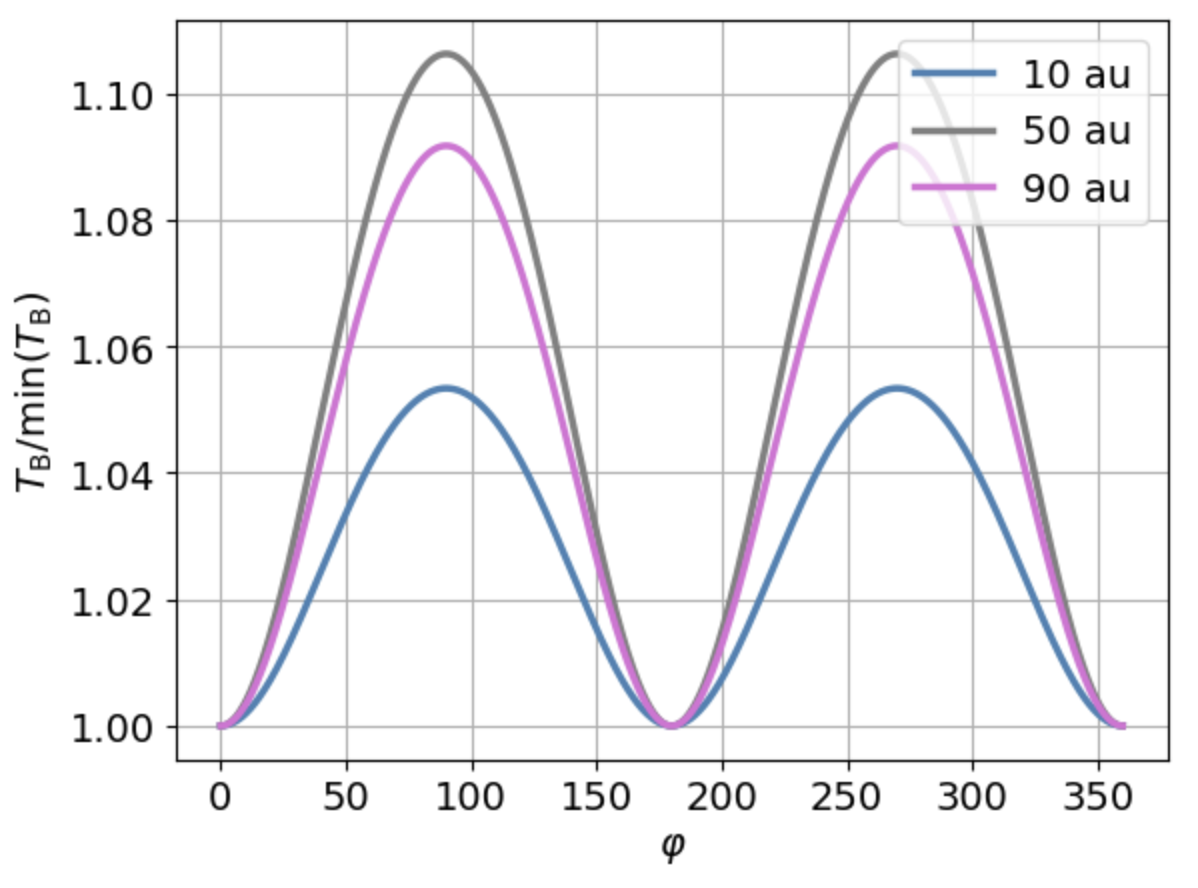}
\caption{{Left panel: optical depth of the background (pink) and the filaments (blue) as modeled. Right panel: normalized brightness temperature at 10 au (blue), 50 au (gray), 90 au (pink).}}
\label{fig:radVariation}
\end{figure*}

{As discussed in \sectionautorefname~\ref{sec:SIsims}, the brightness temperature variation is stronger at intermediate radii because the contrast between the background and filaments' optical depth is stronger.  We illustrate the optical depth radial variation in the left panel of \figureautorefname~\ref{fig:radVariation} for our toy model using typical disc parameters. At small radii, the background is characterized by $\tau_{\rm bg}>1$, making thus the $T_{\rm B}$ oscillation is weak. At intermediate radii, the contrast between $\tau_{\rm bg}$ and $\tau_{\rm c}$ is the strongest, generating the strongest $T_{\rm B}$ oscillation. At large radii, the contrast between $\tau_{\rm bg}$ and $\tau_{\rm c}$ reduces (due to the non-linear relation between the opacity and the grain size), reducing the $T_{\rm B}$ oscillation. We illustrate the normalized $T_{\rm B}$ azimuthal profiles for 3 selected radii in the right panel of \figureautorefname~\ref{fig:radVariation}, confirming that the strongest oscillation is obtained at intermediate radii (50 au).}
 
\bibliography{sample62} 

\begin{thebibliography}{}
\expandafter\ifx\csname natexlab\endcsname\relax\def\natexlab#1{#1}\fi
\providecommand{\url}[1]{\href{#1}{#1}}

\bibitem[{{Adachi} {et~al.}(1976){Adachi}, {Hayashi}, \& {Nakazawa}}]{Adachi+1976}
{Adachi}, I., {Hayashi}, C., \& {Nakazawa}, K. 1976, Progress of Theoretical Physics, 56, 1756

\bibitem[{{Akiyama} {et~al.}(2017){Akiyama}, {Ikeda}, {Pleau}, {Fish}, {Tazaki}, {Kuramochi}, {Broderick}, {Dexter}, {Mo{\'s}cibrodzka}, {Gowanlock}, {Honma}, \& {Doeleman}}]{Akiyama+2017}
{Akiyama}, K., {Ikeda}, S., {Pleau}, M., {et~al.} 2017, \aj, 153, 159

\bibitem[{{Andrews} {et~al.}(2018{\natexlab{a}}){Andrews}, {Terrell}, {Tripathi}, {Ansdell}, {Williams}, \& {Wilner}}]{Andrews+2018b}
{Andrews}, S.~M., {Terrell}, M., {Tripathi}, A., {et~al.} 2018{\natexlab{a}}, \apj, 865, 157

\bibitem[{{Andrews} {et~al.}(2018{\natexlab{b}}){Andrews}, {Huang}, {P{\'e}rez}, {Isella}, {Dullemond}, {Kurtovic}, {Guzm{\'a}n}, {Carpenter}, {Wilner}, {Zhang}, {Zhu}, {Birnstiel}, {Bai}, {Benisty}, {Hughes}, {{\"O}berg}, \& {Ricci}}]{Andrews2018}
{Andrews}, S.~M., {Huang}, J., {P{\'e}rez}, L.~M., {et~al.} 2018{\natexlab{b}}, \apjl, 869, L41

\bibitem[{{Bae} {et~al.}(2023){Bae}, {Isella}, {Zhu}, {Martin}, {Okuzumi}, \& {Suriano}}]{Bae+2023}
{Bae}, J., {Isella}, A., {Zhu}, Z., {et~al.} 2023, in Astronomical Society of the Pacific Conference Series, Vol. 534, Protostars and Planets VII, ed. S.~{Inutsuka}, Y.~{Aikawa}, T.~{Muto}, K.~{Tomida}, \& M.~{Tamura}, 423

\bibitem[{{Bai} \& {Stone}(2010{\natexlab{a}})}]{Bai&Stone2010}
{Bai}, X.-N., \& {Stone}, J.~M. 2010{\natexlab{a}}, \apjl, 722, L220

\bibitem[{{Bai} \& {Stone}(2010{\natexlab{b}})}]{Bai&Stone2010-2}
---. 2010{\natexlab{b}}, \apj, 722, 1437

\bibitem[{{Bai} \& {Stone}(2010{\natexlab{c}})}]{Bai&Stone2010_athenaparticles}
---. 2010{\natexlab{c}}, \apjs, 190, 297

\bibitem[{{Birnstiel} {et~al.}(2018){Birnstiel}, {Dullemond}, {Zhu}, {Andrews}, {Bai}, {Wilner}, {Carpenter}, {Huang}, {Isella}, {Benisty}, {P{\'e}rez}, \& {Zhang}}]{Birnstiel2018}
{Birnstiel}, T., {Dullemond}, C.~P., {Zhu}, Z., {et~al.} 2018, \apjl, 869, L45

\bibitem[{{Bohn} {et~al.}(2022){Bohn}, {Benisty}, {Perraut}, {van der Marel}, {W{\"o}lfer}, {van Dishoeck}, {Facchini}, {Manara}, {Teague}, {Francis}, {Berger}, {Garcia-Lopez}, {Ginski}, {Henning}, {Kenworthy}, {Kraus}, {M{\'e}nard}, {M{\'e}rand}, \& {P{\'e}rez}}]{Bohn+2022}
{Bohn}, A.~J., {Benisty}, M., {Perraut}, K., {et~al.} 2022, \aap, 658, A183

\bibitem[{{Brauer} {et~al.}(2008){Brauer}, {Dullemond}, \& {Henning}}]{Brauer+2008}
{Brauer}, F., {Dullemond}, C.~P., \& {Henning}, T. 2008, \aap, 480, 859

\bibitem[{Brogan {et~al.}(2015)Brogan, Pérez, Hunter, Dent, Hales, Hills, Corder, Fomalont, Vlahakis, Asaki, Barkats, Hirota, Hodge, Impellizzeri, Kneissl, Liuzzo, Lucas, Marcelino, Matsushita, Nakanishi, Phillips, Richards, Toledo, Aladro, Broguiere, Cortes, Cortes, Espada, Galarza, Appadoo, Ramirez, Humphreys, Jung, Kameno, Laing, Leon, Marconi, Mignano, Nikolic, Nyman, Radiszcz, Remijan, Rodón, Sawada, Takahashi, Tilanus, Vilaro, Watson, Wiklind, Akiyama, Chapillon, Monsalvo, Francesco, Gueth, Kawamura, Lee, Luong, Mangum, Pietu, Sanhueza, Saigo, Takakuwa, Ubach, Kempen, Wootten, Carrizo, Francke, Gallardo, Garcia, Gonzalez, Hill, Kaminski, Kurono, Liu, Lopez, Morales, Plarre, Schieven, Testi, Videla, Villard, Andreani, Hibbard, \& Tatematsu}]{ALMA2015}
Brogan, C.~L., Pérez, L.~M., Hunter, T.~R., {et~al.} 2015, The Astrophysical Journal, 808, L3.
\newblock \url{http://dx.doi.org/10.1088/2041-8205/808/1/L3}

\bibitem[{{Carrera} {et~al.}(2015){Carrera}, {Johansen}, \& {Davies}}]{Carrera+2015}
{Carrera}, D., {Johansen}, A., \& {Davies}, M.~B. 2015, \aap, 579, A43

\bibitem[{{Chiang} \& {Youdin}(2010)}]{Chiang&Youdin2010}
{Chiang}, E., \& {Youdin}, A.~N. 2010, Annual Review of Earth and Planetary Sciences, 38, 493

\bibitem[{{Clark}(1980)}]{Clark1980}
{Clark}, B.~G. 1980, \aap, 89, 377

\bibitem[{{Cornwell}(2008)}]{Cornwell2008}
{Cornwell}, T.~J. 2008, IEEE Journal of Selected Topics in Signal Processing, 2, 793

\bibitem[{{D'Alessio} {et~al.}(2004){D'Alessio}, {Calvet}, {Hartmann}, {Muzerolle}, \& {Sitko}}]{D'Alessio+2004}
{D'Alessio}, P., {Calvet}, N., {Hartmann}, L., {Muzerolle}, J., \& {Sitko}, M. 2004, in Star Formation at High Angular Resolution, ed. M.~G. {Burton}, R.~{Jayawardhana}, \& T.~L. {Bourke}, Vol. 221, 403

\bibitem[{{Doi} \& {Kataoka}(2021)}]{Doi&Kataoka2021}
{Doi}, K., \& {Kataoka}, A. 2021, \apj, 912, 164

\bibitem[{{Draine}(2003)}]{Draine2003}
{Draine}, B.~T. 2003, \araa, 41, 241

\bibitem[{{Drazkowska} {et~al.}(2023){Drazkowska}, {Bitsch}, {Lambrechts}, {Mulders}, {Harsono}, {Vazan}, {Liu}, {Ormel}, {Kretke}, \& {Morbidelli}}]{Drazkowska+2023}
{Drazkowska}, J., {Bitsch}, B., {Lambrechts}, M., {et~al.} 2023, in Astronomical Society of the Pacific Conference Series, Vol. 534, Protostars and Planets VII, ed. S.~{Inutsuka}, Y.~{Aikawa}, T.~{Muto}, K.~{Tomida}, \& M.~{Tamura}, 717

\bibitem[{{Facchini} {et~al.}(2020){Facchini}, {Benisty}, {Bae}, {Loomis}, {Perez}, {Ansdell}, {Mayama}, {Pinilla}, {Teague}, {Isella}, \& {Mann}}]{Facchini+2020}
{Facchini}, S., {Benisty}, M., {Bae}, J., {et~al.} 2020, \aap, 639, A121

\bibitem[{{H{\"o}gbom}(1974)}]{Hogbom1974}
{H{\"o}gbom}, J.~A. 1974, \aaps, 15, 417

\bibitem[{{Jennings} {et~al.}(2020){Jennings}, {Booth}, {Tazzari}, {Rosotti}, \& {Clarke}}]{Jennings+2020}
{Jennings}, J., {Booth}, R.~A., {Tazzari}, M., {Rosotti}, G.~P., \& {Clarke}, C.~J. 2020, \mnras, 495, 3209

\bibitem[{{Johansen} {et~al.}(2006){Johansen}, {Klahr}, \& {Henning}}]{Johansen+2006}
{Johansen}, A., {Klahr}, H., \& {Henning}, T. 2006, \apj, 636, 1121

\bibitem[{{Johansen} {et~al.}(2007){Johansen}, {Oishi}, {Mac Low}, {Klahr}, {Henning}, \& {Youdin}}]{Johansen+2007}
{Johansen}, A., {Oishi}, J.~S., {Mac Low}, M.-M., {et~al.} 2007, \nat, 448, 1022

\bibitem[{{Johansen} \& {Youdin}(2007)}]{Johansen&Youdin2007}
{Johansen}, A., \& {Youdin}, A. 2007, \apj, 662, 627

\bibitem[{{Li} {et~al.}(2018){Li}, {Youdin}, \& {Simon}}]{Li+2018}
{Li}, R., {Youdin}, A.~N., \& {Simon}, J.~B. 2018, \apj, 862, 14

\bibitem[{{Min} {et~al.}(2017){Min}, {Stolker}, {Dominik}, \& {Benisty}}]{Min+2017}
{Min}, M., {Stolker}, T., {Dominik}, C., \& {Benisty}, M. 2017, \aap, 604, L10

\bibitem[{{Nakazato} {et~al.}(2019){Nakazato}, {Ikeda}, {Akiyama}, {Kosugi}, {Yamaguchi}, \& {Honma}}]{Nakazato+2019}
{Nakazato}, T., {Ikeda}, S., {Akiyama}, K., {et~al.} 2019, in Astronomical Society of the Pacific Conference Series, Vol. 523, Astronomical Data Analysis Software and Systems XXVII, ed. P.~J. {Teuben}, M.~W. {Pound}, B.~A. {Thomas}, \& E.~M. {Warner}, 143

\bibitem[{{Perkins} {et~al.}(2015){Perkins}, {Marais}, {Zwart}, {Natarajan}, {Tasse}, \& {Smirnov}}]{Perkins+2015}
{Perkins}, S.~J., {Marais}, P.~C., {Zwart}, J.~T.~L., {et~al.} 2015, Astronomy and Computing, 12, 73

\bibitem[{{Pinte} \& {Laibe}(2014)}]{Pinte&Laibe2014}
{Pinte}, C., \& {Laibe}, G. 2014, \aap, 565, A129

\bibitem[{{Scardoni} {et~al.}(2021){Scardoni}, {Booth}, \& {Clarke}}]{Scardoni+2021}
{Scardoni}, C.~E., {Booth}, R.~A., \& {Clarke}, C.~J. 2021, \mnras, 504, 1495

\bibitem[{{Takeuchi} \& {Lin}(2002)}]{Takeuchi&Lin2002}
{Takeuchi}, T., \& {Lin}, D.~N.~C. 2002, \apj, 581, 1344

\bibitem[{{Tazzari} {et~al.}(2018){Tazzari}, {Beaujean}, \& {Testi}}]{Tazzari+2018}
{Tazzari}, M., {Beaujean}, F., \& {Testi}, L. 2018, \mnras, 476, 4527

\bibitem[{{Tazzari} {et~al.}(2020{\natexlab{a}}){Tazzari}, {Clarke}, {Testi}, {Williams}, {Facchini}, {Manara}, {Natta}, \& {Rosotti}}]{Tazzari+2020b}
{Tazzari}, M., {Clarke}, C.~J., {Testi}, L., {et~al.} 2020{\natexlab{a}}, arXiv e-prints, arXiv:2010.02249

\bibitem[{{Tazzari} {et~al.}(2020{\natexlab{b}}){Tazzari}, {Testi}, {Natta}, {Williams}, {Ansdell}, {Carpenter}, {Facchini}, {Guidi}, {Hogherheijde}, {Manara}, {Miotello}, \& {van der Marel}}]{Tazzari+2020a}
{Tazzari}, M., {Testi}, L., {Natta}, A., {et~al.} 2020{\natexlab{b}}, arXiv e-prints, arXiv:2010.02248

\bibitem[{{Tripathi} {et~al.}(2017){Tripathi}, {Andrews}, {Birnstiel}, \& {Wilner}}]{Tripathi+2017}
{Tripathi}, A., {Andrews}, S.~M., {Birnstiel}, T., \& {Wilner}, D.~J. 2017, \apj, 845, 44

\bibitem[{{Warren} \& {Brandt}(2008)}]{Warren&Brandt2008}
{Warren}, S.~G., \& {Brandt}, R.~E. 2008, Journal of Geophysical Research (Atmospheres), 113, D14220

\bibitem[{{Weidenschilling}(1977)}]{Weidenschilling1977}
{Weidenschilling}, S.~J. 1977, \mnras, 180, 57

\bibitem[{{Yang} \& {Johansen}(2014)}]{Yang&Johansen2014}
{Yang}, C.-C., \& {Johansen}, A. 2014, \apj, 792, 86

\bibitem[{{Yang} {et~al.}(2017){Yang}, {Johansen}, \& {Carrera}}]{Yang+2017}
{Yang}, C.~C., {Johansen}, A., \& {Carrera}, D. 2017, \aap, 606, A80

\bibitem[{{Youdin} \& {Johansen}(2007)}]{Youdin&Johansen2007}
{Youdin}, A., \& {Johansen}, A. 2007, \apj, 662, 613

\bibitem[{{Youdin} \& {Goodman}(2005)}]{Youdin&Goodman2005}
{Youdin}, A.~N., \& {Goodman}, J. 2005, \apj, 620, 459

\bibitem[{{Zawadzki} {et~al.}(2023){Zawadzki}, {Czekala}, {Loomis}, {Quinn}, {Grzybowski}, {Frazier}, {Jennings}, {Nizam}, \& {Jian}}]{Zawadzki+2023}
{Zawadzki}, B., {Czekala}, I., {Loomis}, R.~A., {et~al.} 2023, \pasp, 135, 064503

\bibitem[{{Zhu} {et~al.}(2019){Zhu}, {Zhang}, {Jiang}, {Kataoka}, {Birnstiel}, {Dullemond}, {Andrews}, {Huang}, {P{\'e}rez}, {Carpenter}, {Bai}, {Wilner}, \& {Ricci}}]{Zhu+2019}
{Zhu}, Z., {Zhang}, S., {Jiang}, Y.-F., {et~al.} 2019, \apjl, 877, L18

\bibitem[{{Zubko} {et~al.}(1996){Zubko}, {Mennella}, {Colangeli}, \& {Bussoletti}}]{Zubko+1996}
{Zubko}, V.~G., {Mennella}, V., {Colangeli}, L., \& {Bussoletti}, E. 1996, in The Role of Dust in the Formation of Stars, ed. H.~U. {K{\"a}ufl} \& R.~{Siebenmorgen}, 333

\end{thebibliography}




\end{document}